# Investigation of Turbulent transition in plane Couette flows Using Energy Gradient Method

Hua-Shu Dou[1*], Boo Cheong Khoo[2]

[1]Temasek Laboratories,
National University of Singapore, Singapore 117411
[2]Department of Mechanical Engineering
National University of Singapore, Singapore 119260
*Email: tsldh@nus.edu.sg; huashudou@yahoo.com

**Abstract**

The energy gradient method has been proposed with the aim of better understanding the mechanism of flow transition from laminar flow to turbulent flow. In this method, it is demonstrated that the transition to turbulence depends on the relative magnitudes of the transverse gradient of the total mechanical energy which amplifies the disturbance and the energy loss from viscous friction which damps the disturbance, for given imposed disturbance. For a given flow geometry and fluid properties, when the maximum of the function $K$ (a function standing for the ratio of the gradient of total mechanical energy in the transverse direction to the rate of energy loss due to viscous friction in the streamwise direction) in the flow field is larger than a certain critical value, it is expected that instability would occur for some initial disturbances. In this paper, using the energy gradient analysis, the equation for calculating the energy gradient function $K$ for plane Couette flow is derived. The result indicates that $K$ reaches the maximum at the moving walls. Thus, the fluid layer near the moving wall is the most dangerous position to generate initial oscillation at sufficient high Re for given same level of normalized perturbation in the domain. The critical value of $K$ at turbulent transition, which is observed from experiments, is about 370 for plane Couette flow when two walls move in opposite directions (anti-symmetry). This value is about the same as that for plane Poiseuille flow and pipe Poiseuille flow (385-389). Therefore, it is concluded that the critical value of $K$ at turbulent transition is about 370-389 for wall-bounded parallel shear flows which include both pressure (symmetrical case) and shear driven flows (anti-symmetrical case).

# 1. Introduction

Although more than a century has passed since the pioneering work of Reynolds (1883) was done, flow transition from laminar flow to turbulence is still not completely understood [1-5]. In practice, the understanding of turbulence transition and generation has great significance for basic sciences and many engineering fields. This issue is intricately related to the instability problem of the base flow subjected to some imposed disturbances [1-2].

In the past, several stability theories have been developed to describe the mechanism of flow instability. These are: (1) The linear stability theory, which goes back to Rayleigh (1880), is a widely used method and has been applied to several problems [6]. For Taylor-Couette flow and Rayleigh-Bernard convective problem, it agrees well with experimental data. However, this theory fails when used for wall-bounded parallel flows such as plane Couette flow, plane Poiseuille flow and pipe Poiseuille flow. (2) The energy method (Orr, 1907) which is based on the Reynolds-Orr equation is another mature method for estimating flow instability [7]. However, agreement could not be obtained between the theoretical predictions and the experiment data. (3) The weakly nonlinear stability theory (Stuart, 1971) emerged in 1960's and the application is very limited (see [8]). (4) The secondary instability theory (Herbert et al, 1988), which was developed most recently, explains some of flow transition phenomena (mainly for the boundary layer flow) better than the other earlier theories (see [9]). However, there are still significant discrepancies between the predictions obtained using this method and experimental data; particularly at transition.

Studies for parallel flows have attracted many scientists with great concern. For these parallel flows, it is observed from experiments that there is a critical Reynolds number $\text{Re}_c$ below which no turbulence can be sustained regardless of the level of imposed disturbance. For the pipe Poiseuille flow, this critical value of Reynolds number is about 2000 from experiments [10,11]. Above this $\text{Re}_c$, the transition to turbulence depends to a large extent on the initial disturbance to the flow. For example, experiments showed that if the disturbances in a laminar flow can be carefully avoided or considerably reduced, the onset of turbulence can be delayed to Reynolds number up to $\text{Re}=O(10^5)$ [12]. Experiments also showed that for $\text{Re} > \text{Re}_c$, only when a threshold of disturbance amplitude is reached, can the flow transition to turbulence occur [13]. Trefethen et al. suggested that the critical amplitude of the disturbance leading to transition varies broadly with the Reynolds number and is associated with an exponent rule of the form, $A \propto \text{Re}^\gamma$ [12]. The magnitude of this exponent has significant implication for turbulence research [12]. Chapman, through a formal asymptotic analysis of the Navier-Stokes equations (for $Re \to \infty$), found $\gamma = -3/2$ and -5/4 for plane Poiseuille flow with streamwise mode and oblique mode, respectively, with generating a secondary instability, and $\gamma = -1$ for plane Couette flow with above both modes. He also examined the boot-strapping route to transition without needing to generate a secondary instability, and found $\gamma = -1$ for both plane Poiseuille flow and plane Couette flow [14]. Recently, Hof et al. [15], used pulsed injection disturbances in experiments, to obtain the normalized disturbance flow rate in the pipe for the turbulent transition, and found it to be inversely proportional to the Re number, i.e., $\gamma = -1$. This experimental result means that the product of the amplitude of the disturbance and the Reynolds number is a constant for the transition to turbulence. This phenomenon must have a physical background, and a physical mechanism for this result was proposed by Dou [16,17].

Dou [16,17] proposed an energy gradient theory with the aim of clarifying the mechanism of flow transition from laminar flow to turbulence. He gave detailed derivations for this method based on Newton's mechanics, and thus it is compatible to Navier-Stokes equations. For plane Poiseuille flow and Hagen-Poiseuille flow, this method yields consistent results with experimental data. This method is also used to explain the mechanism of instability of inflectional velocity profile for viscous flow and this inflectional instability is only valid for pressure driven flows (it should be noticed that inflectional instability is not suitable for plane Couette flow). However, for shear driven flows such as plane Couette flow, the situation is changed since the energy loss could not be obtained directly from Navier-Stokes equations. It should be mentioned that the energy gradient method is a semi-empirical theory based on physical analysis since the critical value of K is observed experimentally and cannot be directly calculated from the theory so far.

In this paper, the energy gradient method is applied to plane Couette flow (Fig.1). The energy loss along the streamwise direction and the energy gradient along the transverse direction as well as the expression of the energy gradient function are derived. It is shown that the critical value of the energy gradient function determined from experiments is about the same as that for Poiseuille flows. Thus, we verify that the critical value of the energy gradient function, determined from experiment, is the same for all wall bounded parallel flows with symmetrical case (pipe Poiseuille flow and plane Poiseuille flow) or anti-symmetrical case (plane Couette flow).

## 2. Energy Gradient Method

Dou [16,17] proposed a mechanism with the aim to clarify the phenomenon of transition from laminar flow to turbulence for wall-bounded shear flows. In this mechanism, the whole flow field is treated as an energy field. It is suggested that the gradient of total mechanical energy in the transverse direction of the main flow and the loss of the total mechanical energy from viscous friction in the streamwise direction dominate the instability phenomena and hence the flow transition for a given disturbance. It is suggested that the energy gradient in the transverse direction has the potential to amplify a velocity disturbance, while the viscous friction loss in the streamwise direction can resist and absorb this disturbance. The flow instability or the transition to turbulence depends on the relative magnitude of these two roles of energy gradient amplification and viscous friction damping of the initial disturbance. The analysis has obtained very consistent agreement for plane Poiseuille flow and pipe Poiseuille flow for Newtonian fluid at the critical condition [16]. It is also demonstrated that an inflection point existence on the velocity profile is a sufficient condition, but not only a necessary condition, for flow instability, for both inviscid and viscous flows. Later, Dou carried out more detailed derivations from physics to give a solid foundation for this model, and explained recent experimental results on the scaling of the threshold of disturbance amplitude with the Reynolds number found in the literature [17]. This method is named the "*energy gradient method.*" Here, we give a short discussion for a better understanding of the work presented in this study.

For a given base flow, the fluid particles may move in an oscillatory fashion along the streamwise direction if they are subjected to a disturbance. With the oscillatory motion, the fluid particle may gain energy ($\Delta E$) via the disturbance, and simultaneously this particle may have energy loss ($\Delta H$) due to the fluid viscosity along the streamwise direction. The following analysis suggests that the magnitudes of $\Delta E$ and $\Delta H$ determine the stability of the flow of fluid particles. For parallel flows, the relative magnitude of the energy gained from the disturbance and the energy loss due to viscous friction determines the disturbance amplification or decay. Thus, for a given flow, a stability criterion can be written as below for a half-period,

In the energy gradient method, it is indicated that the relative magnitude of the energy of fluid particles gained and the energy loss due to viscous friction in a disturbance cycle determines the disturbance amplification or decay. For a given flow, a stability criterion is written as below for the half-period [17],

$$F = \frac{\Delta E}{\Delta H} = \left(\frac{\partial E}{\partial n}\frac{2A}{\pi}\right) \bigg/ \left(\frac{\partial H}{\partial s}\frac{\pi}{\omega}u\right) = \frac{2}{\pi^2}K\frac{A\omega}{u} = \frac{2}{\pi^2}K\frac{v'_m}{u} < Const, \quad (1)$$

and

$$K = \frac{\partial E/\partial n}{\partial H/\partial s}. \quad (2)$$

Here, $F$ is a function of space which expresses the ratio of the energy gained in a half-period by the particle ($\Delta E$) and the energy loss due to viscosity in the half-period ($\Delta H$). $K$ is a dimensionless field variable (function) and expresses the ratio of transversal energy gradient and the rate of the energy loss along the streamline, which can be calculated from Navier-Stokes equations. $E = \frac{1}{2}\rho V^2$ is the kinetic energy per *unit volumetric fluid*, $s$ is along the streamwise direction and $n$ is along the transverse direction. $H$ is the energy loss per *unit volumetric fluid* along the streamline for finite length. Furthermore, ρ is the fluid density, and u is the streamwise velocity of the main flow. Here, $v'_m = A\omega$ is the disturbance amplitude of velocity and the disturbance has a period of $T = 2\pi/\omega$, $A$ is the amplitude of disturbance in the transverse direction, and $\omega$ is the frequency of the disturbance.

In Eq.(1), when $\Delta E$ is large and $\Delta H$ is small, F will be very large. When F reaches a magnitude larger than a critical value, the flow will be unstable. Otherwise, the flow is stable and keeps laminar. Therefore, it can be found from Eq.(1) that the instability of a flow depends on the values of $K$ and the amplitude of the relative disturbance velocity $v'_m/u$. For all types of flows, it has been shown that the magnitude of $K$ is proportional to the global Reynolds number ($\text{Re} = \rho UL/\mu$) for a given geometry [16]. Thus, the criterion of Eq.(1) can be written as,

$$\text{Re}\frac{v'_m}{u} < Const. \quad (3)$$

For a given flow geometry, U is a characteristic velocity and generally a function of u. Thus, Eq.(3) can be written as,

$$\text{Re}\frac{v'_m}{U} < Const, \quad (4)$$

or

$$(\frac{v'_m}{U})_c \sim (\text{Re})^{-1}. \quad (5)$$

This scaling has been confirmed by careful experiments for the pipe flow [15]. For plane Couette flow, although there is no experimental data available for this scaling, this result of exponent explains the results by Chapman with a formal asymptotic analysis of the Navier-Stokes equations (for Re $\to \infty$ ), for both streamwise mode and oblique mode, with generating a secondary instability. Chapman also examined the boot-strapping route to transition without needing to generate a secondary instability, and found that $\gamma = -1$ for both plane Poiseuille flow and plane Couette flow [14].

The maximum of $F$ in the field will reach its critical value with the increase of Re (see Eq.(1)). The critical value of $F$ indicates the onset of instability in the flow at this location and the initiation of flow *transition* to turbulence. Therefore, at the onset of turbulence, the transition from laminar to turbulent flows is localized. Experiment confirmed that the turbulent spot is actually a localized phenomenon which results from the hairpin vortices [1]. As observed from experiments, a small region of turbulence is generated in the flow at a relatively low Re number, while the turbulence is generated in the full domain at a high Re [1].

In the energy gradient method [16,17], it is seen that the instability of the flow depends on the relative magnitude of the energy gradients in the transverse direction and streamwise direction, for a given disturbance. It is found that the gradient of the total mechanical energy in the transverse direction has a potential to amplify a velocity disturbance, while the viscous friction loss in the streamwise direction can resist and absorb this disturbance energy. The transition to turbulence therefore depends on the relative magnitude of the two roles of energy gradient amplification and viscous friction damping to the initial disturbance. The parameter $K$ as defined in Eq.(1) is a field variable. Thus, the distribution of $K$ in the flow field and the property of disturbance may be the perfect means to describe the disturbance amplification or decay in the flow. We suggested that the flow instability will first occur at the position of $K_{\max}$ which is construed to be the most "dangerous" position. Thus, for a given disturbance, the occurrence of instability depends on the magnitude of this dimensionless function $K$ and the critical condition is determined by the maximum value of $K$ in the flow. For a given flow disturbance, there is a critical value of $K_{\max}$ over which the flow becomes unstable. We emphasize that $K_{\max}$ is the maximum of the magnitude of K in the flow domain at a given flow condition and geometry, while $K_c$ is the critical value of $K_{\max}$ for instability initiation for a given geometry. For a given flow geometry and fluid properties, when the maximum of $K$ in the flow field exceeds a critical value $K_c$, it is expected that instability can occur for a certain initial disturbance [16, 17]. Thus, it is known that turbulence transition is a local phenomenon in the earlier stage. For a given flow, K is proportional to the global Reynolds number. A large value of K has the ability to amplify the disturbance, and vice versa. The analysis for Poiseuille flows supported the idea that the transition to turbulence is due to the energy gradient and the disturbance amplification [16,17], rather than a linear instability [12].

For Poiseuille flows, Dou [16, 17] demonstrated that the energy gradient method has led to a consistent value of Kc at the subcritical condition of transition determined by experiments. It is found that Kc=385~389 at the subcritical condition determined by experiments for both plane Poiseuille flow and pipe Poiseuille flow, as pointed out in Table 1. The most unstable position for plane Poiseuille flow and pipe Poiseuille flow occurs at y/h=±0.58 and r/R=0.58, respectively. These said locations have been confirmed by experiments [16, 17]. For plane Poiseuille flow, Nishioka et al [18] did experiment using ribbon induced disturbance on the flow and showed that the averaged velocity profile displays intense oscillation first in the range of y/h=0.50~0.62. For pipe Poiseuille flow, Nishi et al [19] did experiment using normal injection as disturbance and showed that the averaged velocity profile is subjected to intensive oscillation within r/R=0.53~0.73 during the transition occurrence. These locations observed in experiments are in agreement with the prediction of the energy gradient method, say 0.58.

In plane Couette flow, the streamwise energy gradient (energy loss) for unit volumetric fluid could not be obtained directly from the Navier-Stokes equation as for Poiseuille flows since the flow is uniform along the streamwise direction. Using the energy analysis method, the equation for calculating the energy gradient function $K$ in plane Couette flow is derived in the following section.

## 3. Energy Gradient Method Applied to Plane Couette Flow

In plane Couette flow, the viscous term $\mu\nabla^2\mathbf{u}$ in Navier-Stokes equation is zero, and the total mechanical energy $p+\frac{1}{2}\rho V^2$ per unit volume is constant along the streamwise direction. This is not to say that there is no energy loss due to friction in the flow. Friction loss must still occur since this is a viscous flow (Zero energy loss only occurs in inviscid flow). The energy magnitude is kept constant because the energy loss due to viscous friction is exactly compensated by the energy input to the flow by the moving walls. The work done on the flow by the wall is balanced by the energy loss in the flow.

The velocity distribution for plane Couette flow can be obtained by solving the Navier-Stokes equation and applying the boundary conditions, as in [1, 3].

$$u = ky = \frac{U}{h} y. \qquad (6)$$

Here, $k = \partial u / \partial y$ is the shear rate and is determined by $k = U/h$. The velocity profile is shown in Fig.1. The shear stress is calculated as

$$\tau = \mu \partial u / \partial y = \mu k. \qquad (7)$$

The energy gradient in the transverse direction is calculated by,

$$\frac{dE}{dy} = \rho V \frac{\partial V}{\partial y} = \rho k y \cdot k = \rho k^2 y. \qquad (8)$$

Taking an element in the fluid layer as shown in Fig.2, the work done to the fluid element of length of $\Delta x$ by the upper layer is

$$A_1 = \tau \cdot \Delta x \cdot \Delta z \cdot (u + \Delta u) dt.$$

The work done on the lower layer by the fluid element is

$$A_2 = \tau \cdot \Delta x \cdot \Delta z \cdot u dt.$$

Therefore, the net work done on the fluid element in time $dt$ is given as

$$\Delta A = A_1 - A_2 = \tau \cdot \Delta x \cdot \Delta z \cdot \Delta u dt.$$

The fluid volume passing through $dy$ depth in time $dt$ is

$$\Delta Q = \Delta y \cdot \Delta z \cdot u dt.$$

Hence, the energy consumed by the element per unit volume of fluid over the length of $\Delta x$ (Fig.2) is

$$\Delta H = \frac{\Delta A}{\Delta Q} = \frac{\tau \Delta x \Delta z \Delta V dt}{\Delta y \Delta z V dt} = \frac{\tau \Delta x \Delta u}{\Delta y \cdot u}. \tag{9}$$

The energy loss of unit volume of fluid in the length of $\Delta x$ is equal to the energy consumed since the energy is constant along the streamline. Then, the energy loss per volumetric fluid per unit length along the streamwise direction is given by

$$\frac{\Delta H}{\Delta x} = \frac{\tau}{u}\frac{\Delta u}{\Delta y}. \tag{10}$$

Thus, the rate of energy loss along the streamline direction from Eq.(10) is

$$\frac{dH}{dx} \equiv \frac{\tau}{u}\frac{du}{dy}. \tag{11}$$

Substituting Eq.(8) and (11) into Eq.(2), the ratio of the energy gradient in transverse direction and the energy loss in streamwise direction can be written as,

$$K = \frac{\rho u(\partial u/\partial y)}{dH/dx} = \frac{\rho k k y}{\frac{\tau}{u}\frac{du}{dy}} = \frac{\rho U h}{\mu}\frac{y^2}{h^2} = \mathrm{Re}\frac{y^2}{h^2}, \tag{12}$$

where $\mathrm{Re} = \rho U h/\mu$ is the Reynolds number. It can be seen that the magnitude of K is proportional to Re at any location in the flow field. K is a quadratic function of y/h across the channel width. There is no maximum within the channel unlike for Poiseuille flows [16]. It reaches its maximum only on the walls (y=±h),

$$K_{\max} = \frac{\rho U h}{\mu} = \mathrm{Re}. \tag{13}$$

The variations $dE/dy$, $dH/dx$, and K along the channel width are shown in Figs.3-5 for plane Couette flows with the two plates moving in opposite directions. It can be seen from Fig.3 that the gradient of the mechanical energy along the transverse direction in plane Couette flow is linear along the width of the channel and it attains its maximum at the walls. Therefore, the fluid layer near the wall has the greatest potential to amplify a disturbance. From Fig.4, it is observed that the energy loss distribution in plane Couette flow is lower at the walls and it attains its largest value at the centre of the channel. Therefore, the ability of flow to damp a disturbance is low near the walls and is large at the centre of the channel. The mechanism of energy loss damping disturbance has been detailed in [20] for both plane Couette flow and Taylor-Couette flow. It is found from Fig. 5 that value of K is zero at the centreline and it reaches its maximum at the walls. From the energy gradient method [16, 17], Eq.(1) indicates that $K(v'_m/u) < Const$ is the criteria for stability for parallel flows. Thus, the flow is expected to be more unstable where $K$ is higher than that where K is lower, for given same level of normalized perturbation $(v'_m/u)$. The first instability should be associated with the maximum of K, $K_{max}$, in the flow field for same given disturbance level. Therefore, the flow near the wall is the most dangerous position to amplify a given disturbance for plane Couette flow if the disturbance is uniformly distributed.

## 4. Discussions

### 4.1 Instability mechanism and disturbance amplification

As is well known, the development of a disturbance is subjected to the governing equations (i.e., Navier-Stokes equations) and the boundary and initial conditions. The value of K represents the effect of the governing equations on the disturbance. Thus, the flow stability depends on the distribution of K in the flow field and the initial disturbance provided to the flow. On the other hand, we should distinguish between the disturbance in laminar state and the velocity fluctuation in turbulent state. The laminar flow is completely different from turbulent flow regarding the disturbance. The place where the disturbance is the largest in laminar flow is not necessarily the same as that where the turbulent stresses are largest in the corresponding turbulent state. Near the moving wall, the capacity of the base flow to amplify a disturbance is largest owing to the maximal magnitude of K (Fig.5). However, actually, the flow disturbances at the wall should be vanishing due to no-slip condition. Therefore, it is likely that the most dangerous position is not directly at the wall, but at a location (very) near the wall where an initial disturbance is present and yet K has a large magnitude. Thus, a small disturbance could be easily amplified by the large energy gradient at such position. Therefore, the fluid layer near the moving wall is the most dangerous position to generate initial oscillation. This mechanism is obviously correct from the principle of energy conservation. *Turbulence sustenance is maintained by input of energy from* *external sources, otherwise it would die.* The moving wall is the object to put the energy into the flow and therefore it has the power to sustain the turbulence. The place where the energy is higher should be the region of intense turbulent fluctuation.

Figure 6 shows the measurements by Bottin et al [21] for plane Couette flow during the process leading to the formation of a turbulent spot. Three slices of profiles are sketched within a turbulent spot in plane Couette flow. This picture was taken for the flow near the critical condition (Re=340). The profile on the right side is at the edge of the spot and is in the initial instability stage. It is seen that the flow oscillation first occurs near one of the moving walls. From this figure, it is also observed that the process of flow transition is not symmetrical relative to the channel width which might be subjected to the influence of experiment and facility uncertainties. This experiment indicates that the role of the $K_{max}$ dominates near the moving wall, and further lends credibility to the energy gradient theory. Analytical study of Lessen and Cheifetz [22] also showed that the instability of the base flow first starts from the place near one of the walls.

| Flow type | Re expression | Linear stability analysis, $Re_c$ | Energy method $Re_c$ | Experiments, $Re_c$ | Energy gradient method, $K_{max}$ at $Re_c$ (from experiments), $\equiv K_c$ |
|---|---|---|---|---|---|
| Pipe Poiseuille | $Re = \rho UD/\mu$ | Stable for all Re | 81.5 | 2000 | 385 |
| Plane Poiseuille | $Re = \rho UL/\mu$ | 7696 | 68.7 | 1350 | 389 |
|  | $Re = \rho u_0 h/\mu$ | 5772 | 49.6 | 1012 | 389 |
| Plane Couette | $Re = \rho Uh/\mu$ | Stable for all Re | 20.7 | 370 | 370 |

Table 1 Comparison of the critical Reynolds number and the energy gradient parameter $K_{max}$ for plane Poiseuille flow and pipe Poiseuille flow as well as for plane Couette flow [16, 17]. *U* is the averaged velocity, $u_0$ the velocity at the mid-plane of the channel, *D* the diameter of the pipe, *h* the half-width of the channel for plane Poiseuille flow (L=2h) and plane Couette flow. The experimental data for plane Poiseuille flow and pipe Poiseuille flow are taken from Patel and Head [11]. The experimental data for plane Couette flow is taken from Tillmark and Alfredsson [27], Daviaud et al [28], and Malerud et al [29]. Here, two Reynolds numbers are used since both definitions are employed in literature. The data of critical Reynolds number from energy method are taken from [1]. For Plane Poiseuille flow and pipe Poiseuille flow, the $K_{max}$ occurs at y/h=0.58, and r/R=0.58, respectively. For plane Couette flow, the $K_{max}$ occurs at y/h=1.0.

**4.2 Comparison with experiments at critical condition**

Earlier experimental results on plane Couette flow showed that the critical Reynolds number lies in the range of 280 to 750 [23-24]. Lundbladh and Johansson (1991)'s direct numerical simulation produced a critical condition of Rec=375 for plane Couette flow [25]. For this type of flow, Hegseth et al [26] observed an intermittent turbulent state which occurs in the range of Re of 380--450. Below this range, the laminar state is stable to finite amplitude perturbation and above this range the entire flow domain is turbulent. For 380<Re<450 and after a perturbation is imposed, the dynamical regime shows a fluctuating mixture of laminar and turbulent domains which is reminiscent of spatiotemporal intermittency. This result showed that the minimum Re for the transition with finite amplitude disturbance is about 380. Tillmark and Alfredsson [27], Davidud et al [28], and Malerud et al [29] carried out some experiments for turbulent transition for plane Couette flows using flow visualization techniques. All of these experiments showed that the critical condition occurs at about Rec=370 ± 10. Although the subsequent experimental results showed a lower critical Reynolds number (325~380) [30-31], this does not detract from the comparisons carried out here. Using the experimental data Rec=370, we obtain Kc=370 from Eq.(13) at the critical condition determined by experiments below which no turbulence occurs (see Table 1). This critical value of Kc=370 is near to the value for Poiseuille flows, 385~389. The small difference in the value obtained is subjected to the uncertainty of the critical condition in experiments. For example, the determination of transition is deduced from the abrupt change in the drag coefficient as found by Patel and Head [11], while the flow visualization method is used in [30-31]. These results demonstrate that the critical value of $K_{max}$ for wall-bounded parallel flows including both pressure driven and shear driven flows is about 370~389. This consistency also suggests that the mechanisms of instabilities in wall-bounded parallel shear flow are perhaps the same. They are all dominated by the transverse energy gradient and the streamwise flow energy loss. The results obtained in this study provide further basis for better understanding of the mechanism of instability and transition to turbulence in parallel shear flows.

**4.3 Flux of vorticity rather than vorticity to dominate transition**

In turbulence modelling, the turbulent stress is generally modelled in terms of the velocity gradient of mean flows or the strain-rate-tensor. The magnitude of the velocity gradient or the vorticity is an indication of the strength of turbulent stress in some cases. Thus, it may be deduced that the turbulent transition might be related to the magnitude of the velocity gradient. However, from the energy gradient method, the magnitude of the velocity gradient or vorticity is not the dominating factor influencing the transition. Instead, it is the flux of vorticity ($u\partial u/\partial y$ in 2D parallel flow case) or the energy gradient which is the governing factor. Figure 7 depicts the development of plane Couette flow with the increase of the channel width. In these three cases, the shear rate (velocity gradient) is kept the same and the width of the channel is allowed to vary. With increasing channel width, Re increases. As we could see from experiments, the flow remained laminar in (a) and (b) when the Re is lower than the critical Reynolds number. The flow in (c) becomes turbulent when Re is higher than a critical value. It is therefore suggested that the transition to turbulence is not due directly to the influence of the velocity gradient although it depends on Re. From Fig.7 and Eq.(8), it is seen that the maximum energy gradient (or flux of vorticity) which is located close to the walls increases with the channel width (and also Re). From Fig.7 and Eq.(11), it is also seen that the energy loss near the walls decreases with the channel width (and also Re). Therefore, the turbulent transition for large width is attributed to the increase of the transverse energy gradient (or flux of vorticity in 2D parallel flow) near the walls and the decrease of the energy loss in the streamwise direction near the walls, rather than the magnitude of velocity gradient or vorticity.

## 5. Summary and Conclusion

In this paper, the instability of plane Couette flow is studied using the energy gradient method. The expression of K, the ratio of the gradient of the total mechanical energy in the transverse direction to the rate of loss of the total mechanical energy in streamwise direction, is derived using energy analysis. It is shown that the transverse gradient of the mechanical energy (or the flux of the vortex) is destabilizing and the energy loss due to viscosity is stabilizing. At the transition, it is found that the critical value of $K_{\max}$ determined from experimental data is about 370 (or 325~370) for plane Couette flow (two plates move in opposite directions) and the most unstable position is near the walls. In the range of Re mentioned above (325~370 or so), it has been found in experiment that there is a behaviour of intermittent of turbulence in which laminar flow and turbulent flow co-exist in the domain [32], which means that this flow range of Re is the transition stage. *It is interesting to note that based on the critical conditions determined by experiments, the critical value of K for plane Couette flow (two plates move in opposite directions) is about the same as that for plane Poiseuille flow and pipe Poiseuille flow.* Therefore, it is suggested that the critical value of *K* at turbulent transition is about 370-389 for wall-bounded parallel shear flows which include both pressure (symmetrical case) and shear driven flows (anti-symmetrical case). These results show that the energy gradient method is universal for both pressure and shear driven flows. In separate works, the energy gradient method is also demonstrated to be valid for Taylor-Couette flow between concentric rotating cylinders [33], boundary layer flows [34], and straight annulus flows [35], and excellent agreements have been achieved with all the available experimental data in literature. In [36], criteria for turbulent transition have been proposed for pressure and shear driven flows, respectively, following the principle of energy gradient theory. With the proposed theory, turbulence can be controlled with manipulation of the distribution of the energy gradient function in the flow field.

Using the energy gradient method, it is found in plane Couette flows that *the fluid layer near the moving walls is the most dangerous position to generate initial oscillation at sufficient high Re for given same level of normalized perturbation in the domain*. This mechanism has been also explained by the principle of energy transmission between fluid layers since turbulence sustenance is maintained by input of energy from external sources.

The energy gradient method employed here may provide a *universal basis* for the modelling and prediction of the transition process. This is especially relevant since shear-driven flows and pressure-driven flows are very different, and yet the mechanism for transition to turbulence is very similar based on the energy gradient theory. The marked difference between the shear driven and pressure driven flows lies in the energy transmission process: in the former, the energy is transported to the wall from the core, but in the latter it is transported to the core from the wall. This difference determines the feature of the most dangerous position in the flow field for the instability. This difference in energy transmission may be relevant to the different turbulent stress distribution near the wall as simulated by Bech et al. [37]. However, even though there is a difference in the behaviour of flow fields, the transition to turbulence occurs at the same value of the energy gradient quantity (Kmax).

**Acknowledgement**
The authors acknowledge the useful discussions with Prof. KS Yeo.

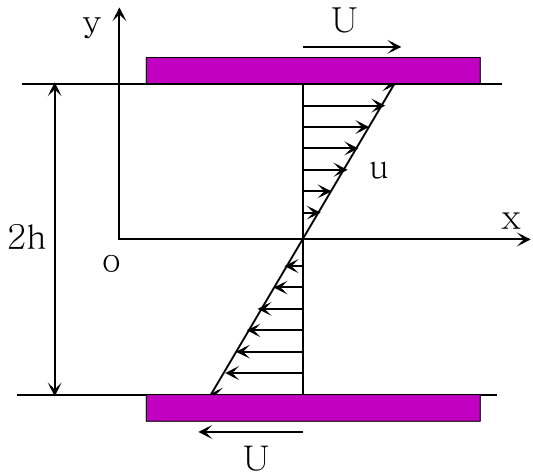

Fig.1 Plane Couette flow with two plates moving in opposite directions (anti-symmetrical case).

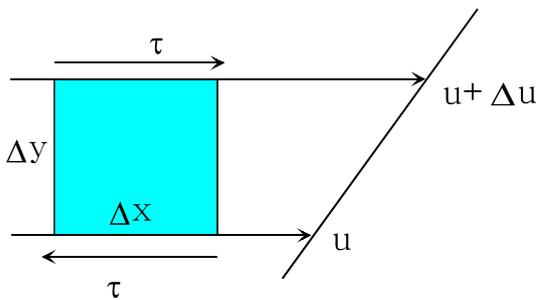

Fig.2 A cubic fluid element in plane Couette flows. $\Delta z$ is perpendicular to x-y plane.

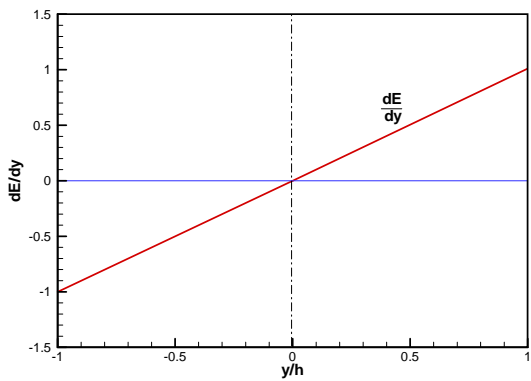

Fig.3 Distribution of energy gradient along the transverse direction in plane Couette flow for anti-symmetrical case shown in Fig.1.

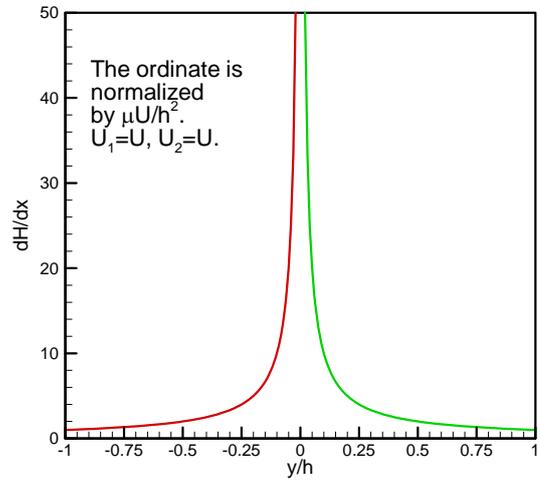

Fig.4 Energy loss distribution in plane Couette flow for anti-symmetrical case shown in Fig.1.

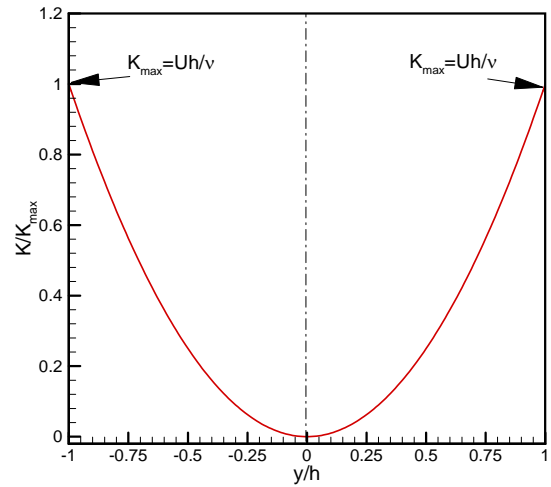

Fig.5 Distribution of $K$ versus the channel width for plane Couette flows with the two plates moving in opposite directions.

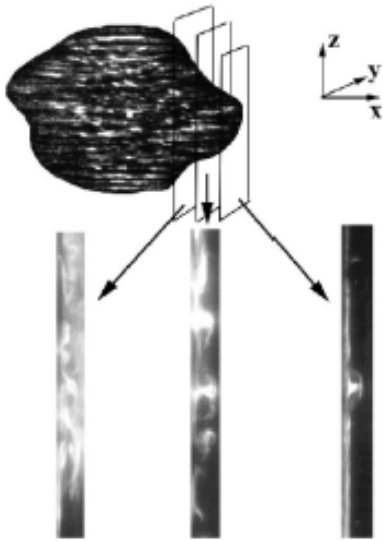

Fig.6 Sectional views of the flow at the border of a turbulent spot in three x=constant planes for Re=340 for plane Couette flow. Here, x is in the streamwise direction, y is in transverse direction, and z is in the spanwise direction (Bottin et al 1998; Courtesy of Dauchot).

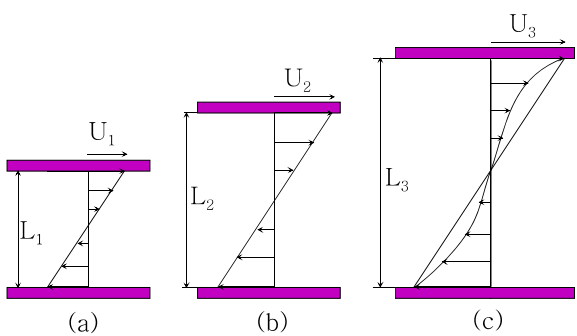

(a)　　　(b)　　　(c)

Fig.7 Development of plane Couette flow with increasing channel width (two plates move in opposite directions; here it is noticed as anti-symmetrical case). The shear rate is the same for three cases. With the increasing of the channel width, Re increases. The flow is laminar for (a) and (b) cases, and it becomes turbulent for (c) with the increasing Re.